\documentclass[prd,twocolumn,nopacs,floatfix,amsmath,nofootinbib,amssymb,floatfix]{revtex4}
\usepackage{graphicx,color,dcolumn,booktabs,bm}
\usepackage{longtable,lscape}
\usepackage{pdfpages}
\usepackage{txfonts}
\usepackage{overpic}
\usepackage{amssymb}
\usepackage{indentfirst}
\usepackage{feynmf}   
\usepackage{slashed}  
\usepackage{cases}
\usepackage{color}
\usepackage{multirow}
\usepackage{threeparttable}
\usepackage{epstopdf}
\usepackage{enumerate}
\usepackage{amsmath}
\usepackage{graphicx,color,dcolumn,booktabs,bm}
\usepackage[colorlinks, citecolor=blue,anchorcolor=red,menucolor=red, linkcolor=red,filecolor=red,urlcolor=blue,frenchlinks=red]{hyperref}

\graphicspath{{Figures/}} %

\begin{document}

\preprint{APS/123-QED}

\title{Production Mechanisms of isoscalar pseudotensor mesons in pion and kaon induced reactions}

\author{Dong-En Lu$^{1}$}
\author{Li-Ming Wang$^{1}$\footnote{Corresponding author}}\email{lmwang@ysu.edu.cn}
\affiliation{ $^1$Key Laboratory for Microstructural Material Physics of Hebei Province, School of Science, Yanshan University, Qinhuangdao 066004, China
}

\date{\today}

\begin{abstract}

The recent observation of the X(2600) resonance by the BESIII Collaboration has motivated a renewed interest in the spectroscopy of light mesons, particularly pseudotensor states. However, a significant theoretical gap exists in the poorly explored spectrum of isoscalar pseudotensor mesons, where several predicted radial excitations remain unobserved. To address this, we investigate the production of four such states ($\eta_2(4D)$, $\eta'_2(2D)$, $\eta'_2(3D)$, and $\eta'_2(4D)$) employing an effective Lagrangian approach combined with Regge trajectory phenomenology to describe the high-energy behavior of the reactions. Our calculations reveal a distinct production selectivity: the $\eta_2(4D)$ state is more prominent in pion induced ($\pi^- p \to \eta_2 n$) reactions, whereas the $\eta'_2(2D, 3D, 4D)$ states are preferentially produced in kaon induced ($K^- p \to \eta_2 \Lambda$) processes. Moreover, the total cross sections for these states exhibit characteristic peaks at specific beam momenta, providing clear kinematic windows for their discovery. This study provides the first comprehensive theoretical predictions for the production cross sections of these predicted mesons, offering crucial guidance for future experimental searches.

\end{abstract}

\maketitle


\section{\label{sec:level1}INTRODUCTION}
The study of light meson spectroscopy provides crucial insights into the nonperturbative regime of Quantum Chromodynamics (QCD). With the world's largest accumulated $J/\psi$ data sample, the BESIII Collaboration has established an excellent platform for probing light hadrons. Their progressively enhanced experimental precision has led to the observation of several resonances, such as $X(1835)$, $X(2120)$, $X(2370)$, and the recently reported $X(2600)$, in the $J/\psi \rightarrow \gamma \pi^{+} \pi^{-} \eta^{\prime}$ decay ~\cite{BES:2005ega,BESIII:2010gmv,BESIII:2022sfx}. Notably, the $X(2600)$ was observed in the $\eta' \pi^+ \pi^-$ invariant mass spectrum with a significance exceeding $20\sigma$, based on approximately $10 \times 10^{10}$ $J/\psi$ events. Its measured mass and width are $2618.3 \pm 2.0\,_{-1.4}^{+16.3}~\text{MeV}/c^2$ and $195 \pm 5\,_{-17}^{+26}~\text{MeV}$, respectively. Although its spin-parity quantum numbers $J^{PC}$ remain undetermined, the observed decay channel suggests possible assignments of $0^{-+}$ (pseudoscalar) or $2^{-+}$ (pseudotensor).

The nature of the $X(2600)$ is actively debated. with interpretations including a $J^{PC} = 2^{-+}$ glueball within QCD sum rules ~\cite{Zhang:2022obn} or a $2^{-+}$ tetraquark state ~\cite{Wang:2025nme}. Furthermore, as discussed in Ref.~\cite{Chen:2022asf}, identifying the $X(2600)$ as a glueball or a conventional state may also depend on the internal structures of the $f_0(1500)$ and $f_2'(1525)$ resonances, adding further complexity to its interpretation.

From a conventional quark-model perspective, the previously observed resonances $X(1835)$, $X(2120)$, and $X(2370)$ in the $\pi^+\pi^-\eta'$ mass spectrum were all identified as $0^{-+}$ pseudoscalar states \cite{Yu:2011ta,Wang:2020due}. This systematic pattern suggests a similar pseudoscalar nature for the $X(2600)$, leading to its interpretation as the $\eta(6S)$ isoscalar pseudoscalar state in Ref.~\cite{Wang:2020due}. However, a pseudotensor assignment cannot be entirely excluded. Crucially, the analysis in Ref.~\cite{Wang:2024yvo} demonstrates that the  decay behaviors disfavor identifying the $X(2600)$ as the $\eta'_2(4D)$ state, effectively ruling out its classification as an isoscalar pseudotensor meson within this framework. This framework predicts four unobserved radial excitations of isoscalar pseudotensor mesons near 2.6 GeV: $\eta_2(\mathrm{4D})$ ($2498$ MeV), $\eta'_2(\mathrm{2D})$ ($2238$ MeV), $\eta'_2(\mathrm{3D})$ ($2520$ MeV), and $\eta'_2(\mathrm{4D})$ ($2764$ MeV). The existence of these predicted states naturally raises the question of how such highly excited pseudotensor mesons might be produced and observed experimentally.

To contextualize our study, we briefly review the known $\eta_2$ states listed by the PDG~\cite{ParticleDataGroup:2024cfk}, which include the $\eta_2(1645)$ \cite{CrystalBarrel:1996bnu,WA102:1997sum}, $\eta_2(1870)$ \cite{WA102:1999ybu,CrystalBarrel:1996bnu,Anisovich:2010nh}, $\eta_2(2030)$ \cite{Anisovich:2000mv}, and $\eta_2(2250)$ \cite{Anisovich:2000us}. These states were primarily observed in $p\bar{p}$ annihilation processes and commonly decay via the $a_2(1320)\pi$ channel. Pion-proton and kaon-proton scattering have long served as effective tools for discovering light hadrons \cite{Wohlmut:1970gp,Briefel:1977bp,Prakhov:2005qb,Shirotori:2012ka,WANG:2025fmh}. For instance, the $\pi_2(1670)$ was first identified in $\pi^+p \to p\pi^+\pi^+\pi^-$ reactions~\cite{Bartsch:1968zz}, and the $K_2(1770)$ was initially observed in $K^-p \to pK^-\pi^+\pi^-$ and $K^-p \to p\bar{K}^0\pi^-\pi^0$ reactions~\cite{Bartsch:1966zz}. As members of the same pseudotensor nonet, the predicted $\eta_2$ states could likewise be produced in similar processes, which forms the primary motivation for the present work.

Building upon this motivation, we investigate the production of the four predicted isoscalar pseudotensor mesons in pion and kaon induced reactions on a proton target, employing an effective Lagrangian approach. A Regge-based phenomenological treatment is applied to describe the high-energy behavior of these processes, successfully reproducing the expected energy and $t$-dependence of differential cross sections in the forward region \cite{Guidal:1997hy}. Given the absence of experimental data for the specific reactions $\pi^-p\to \eta_2n$ and $K^-p\to \eta_2\Lambda$, our study focuses on predicting their energy-dependent cross sections within the Regge framework. These predictions are intended to provide concrete guidance for future experimental searches at facilities such as J-PARC~\cite{Kumano:2015gna}, COMPASS~\cite{Nerling:2012er}, and SPS@CERN~\cite{Velghe:2016jjw}.

This paper is structured as follows. Following this introduction, Sec.~\ref{sec:level2} details the theoretical formalism, including the effective Lagrangians and scattering amplitudes. Sec.~\ref{sec:level3} presents our numerical results and discussion. Finally, the paper concludes with a summary in Sec.~\ref{sec:level4}.

\section{\label{sec:level2}Production Mechanisms for $\eta_2$ Mesons in Pion and Kaon Induced Reactions}

\begin{figure}[h!]
    \centering
    \includegraphics[width=0.45  \textwidth]{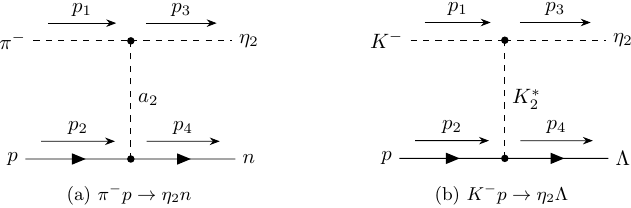} 
    \caption{Feynman diagrams for (a) pion induced and (b) kaon induced production of the \(\eta_2\) meson. The \(t\)-channel exchange is dominated by the \(a_2\) meson in (a) and the \(K_2^*\) meson in (b).}
    \label{fig:1}
\end{figure}

The reaction mechanisms for pion and kaon induced productions of isoscalar pseudotensor mesons are illustrated in the the $t$-channel Feynman diagram of Fig.~\ref{fig:1}. We consider only the $t$-channel contributions, as the $s$-channel is typically negligible and the $u$-channel, which involves the exchange of a nucleon or hyperon, is dominant at backward angles—a region outside the forward-focused scope of this Regge-based analysis.

\subsection{\label{sec:lagrangians}Lagrangians}

The effective Lagrangian densities required for computing the processes in Fig.~\ref{fig:1} are given by \cite{Koenigstein:2016tjw,Yu:2011zu}:
\begin{gather}
\mathcal{L}_{TXP} = g_{TXP} T_{\mu\nu} X^{\mu\nu} P, \\
\mathcal{L}_{TNN} = -2i\frac{g_{TNN}^{(1)}}{M} \bar{N} (\gamma_\lambda \partial_\sigma + \gamma_\sigma \partial_\lambda) N T^{\lambda\sigma} \nonumber \\
\quad + 4\frac{g_{TNN}^{(2)}}{M^2} \partial_\lambda \bar{N} \partial_\sigma N T^{\lambda\sigma},
\end{gather} 
where $T$, $X$, $P$ and $N$ denote the pseudotensor meson, tensor meson, pseudoscalar meson and nucleon fields, respectively.

The coupling constants $g_{TNN}^{(1, 2)}$ are determined using SU(3) flavor symmetry. For the $f_2NN$ vertex, we adopt $g_{f_2NN}^{(1)} = 6.45$ and $g_{f_2NN}^{(2)} = 0$ from Refs.~\cite{Yu:2011fv}, values consistent with SU(3) predictions for parameters $\alpha = 2.0$ and $F/D = -1.8$. The corresponding couplings for $a_2NN$ and $K_2^*N\Lambda$ vertices are determined through the SU(3) relations:
\begin{align}
g_{f_2NN}^{(1,2)} &= \frac{1}{\sqrt{3}} (4\alpha_{(1,2)} - 1) g_{a_2NN}^{(1,2)}, \label{eq:f2nn} \\
g_{K_2^*N\Lambda}^{(1,2)} &= -\frac{1}{\sqrt{3}}(1+2\alpha_{(1,2)})g_{a_2NN}^{(1,2)}, \label{eq:k2nlam}
\end{align}
where we use $g_{a_2NN}^{(1)} = 1.4$ and derive $g_{K_2^*N\Lambda}^{(1)} = -4.45$ from these symmetry relations.

The coupling constants $g_{TXP}$ are determined from the corresponding decay widths, which are expressed as:
\begin{gather}
\Gamma_{T \to X P} = \frac{k_f}{8\pi m_T^2} \frac{g_{TXP}^2}{45} 
\Bigg( 4\frac{k_f^4}{m_X^4} + 30\frac{k_f^2}{m_X^2} + 45 \Bigg). \label{eq:TXP}
\end{gather}
Here, $k_f = k_f(m_T, m_X, m_P)$ is the modulus of the outgoing momentum in the rest frame of $T$, given by:
\begin{equation}
\begin{split}
k_f(m_T, m_X, m_P) = \frac{\lambda^{1/2} ( m_{T}^2, m_{X}^2, m_{P}^2 )}{2m_T}, 
\end{split}
\label{eq:kf}
\end{equation}
where \(\lambda(x, y, z)\) is the Källén function, defined as \(\lambda(x,y,z) = x^2 + y^2 + z^2 - 2xy - 2xz - 2yz\). Utilizing the decay widths provided in Ref.~\cite{Wang:2024yvo}, we calculate the coupling constants $g_{\eta_2 a_2 \pi}$ and $g_{\eta_2 K K_2^{*}}$, which are listed in Table~\ref{tab:eta2_states_properties}.

\begin{table}[h]
\centering
\caption{Coupling constants for $\eta_2$ decays.}
\label{tab:eta2_states_properties}
\renewcommand{\arraystretch}{1.2}
\begin{tabular*}{\linewidth}{@{\extracolsep{\fill}} lcccc}
\hline
& \multicolumn{4}{c}{Pseudotensor State} \\
\cline{2-5}
Parameter & $\eta_2(\mathrm{4D})$ & $\eta_2^{'}(\mathrm{2D})$ & $\eta_2^{'}(\mathrm{3D})$ & $\eta_2^{'}(\mathrm{4D})$ \\
\hline
$g_{\eta_2 a_2\pi}$ (GeV)& 1.104  & 0.955  & 0.698  & 0.566 \\
$g_{\eta_2 K{K_2}^*}$ (GeV) & 0.268  & 2.934  & 2.165  & 1.751  \\
\hline
\end{tabular*}
\end{table}

\subsection{\label{sec:citeref}Amplitudes}
With the above preparation, the amplitudes of these discussed reactions shown in Fig.~\ref{fig:1} can be written as

\begin{gather}
\mathcal{M}_{\pi^{-} p\to\eta_2n} = g_{\eta_2a_2\pi}\epsilon_{\mu\nu}(\eta_2)\frac{\Pi^{\mu\nu;\lambda\sigma}}{t-m_{a_2}^2}\bar{u}(p_4) \nonumber \\
\times\left[\frac{g_{a_2NN}^{(1)}}{M_N}\left(\gamma_\lambda P_\sigma+\gamma_\sigma P_\lambda\right)+\frac{g_{a_2NN}^{(2)}}{M_N^2}P_\lambda P_\sigma\right]u(p_2){F_{a_2}^2(t)},
\label{eq:2}
\end{gather}

\begin{gather}
\mathcal{M}_{K^- p \to\eta_2 \Lambda} = g_{\eta_2 {K_2^*}K}\epsilon_{\mu\nu}(K_2^*)\frac{\Pi^{\mu\nu;\lambda\sigma}}{t-m_{{K_2^*}}^2} \bar{u}_N(p_4) \nonumber \\
\times\left[\frac{g_{{K_2^*}N\Lambda}^{(1)}}{M_N}\left(\gamma_\lambda P_\sigma+\gamma_\sigma P_\lambda\right)+\frac{g_{{K_2^*}N\Lambda}^{(2)}}{M_N^2}P_\lambda P_\sigma\right]u_\Lambda(p_2){F_{{K_2^*}}^2(t)}.
\label{eq:4}
\end{gather}
The tensor propagator structure is given by:
\begin{gather}
\Pi_{\mu\nu;\lambda\sigma}= \frac{1}{2}(\bar{g}_{\mu\lambda}\bar{g}_{\nu\sigma} + \bar{g}_{\mu\sigma}\bar{g}_{\nu\lambda}) - \frac{1}{3}\bar{g}_{\mu\nu}\bar{g}_{\lambda\sigma},
\end{gather}
with
\begin{gather}
\bar{g}_{\mu\nu} = -g_{\mu\nu} + \frac{p_{t\mu} p_{t\nu}}{m_{t}^2}
\end{gather}
For the $t$-channel exchange, we employ a form factor of the type $F_{x}(t) = (\Lambda_x^2 - m_{x}^2)/(\Lambda_x^2 - t)$, where $t = (p_1 - p_3)^2$ is the Mandelstam variable. The cutoff parameter $\Lambda_x$ in the form factor will be discussed in Sec.~\ref{sec:level3}

\subsection{Reggeized $t$-channel}%

To accurately describe the high-energy behavior of the cross section, we incorporate Regge phenomenology into the $t$-channel exchange ~\cite{Nam:2011tj,Wang:2018mjz, Wang:2019qyy, Wang:2023lia}. This is achieved by replacing the standard Feynman propagators
in Eqs.~\eqref{eq:2} and \eqref{eq:4} with the corresponding Reggeized propagators \cite{Yu:2011zu,Yu:2011fv}:
\begin{align}
\frac{1}{t - m^2_{a_2}} &\rightarrow 
\left( \frac{s}{s_{\mathrm{scale}}} \right)^{\alpha_{a_2}(t)-2} 
\frac{\pi \alpha_{a_2}^{\prime}(t)}{\Gamma[\alpha_{a_2}(t)-1] \sin[\pi \alpha_{a_2}(t)]} \nonumber \\
&\quad \times \left( \frac{1 + e^{-i\pi \alpha_{a_2}(t)}}{2} \right), \\
\frac{1}{t - m^2_{K_2^*}} &\rightarrow 
\left( \frac{s}{s_{\mathrm{scale}}} \right)^{\alpha_{K_2^*}(t)-2} 
\frac{\pi \alpha_{K_2^*}^{\prime}(t)}{\Gamma[\alpha_{K_2^*}(t)-1] \sin[\pi \alpha_{K_2^*}(t)]} \nonumber \\
&\quad \times \left( \frac{1 + e^{-i\pi \alpha_{K_2^*}(t)}}{2} \right).
\end{align}
Here $s_{\mathrm{scale}}$ is a mass scale, which is conventionally chosen as $s_{\mathrm{scale}}$=1 GeV$^2$. The Regge trajectories for the $a_2$ and $K_2^*$ exchanges are parametrized as:
\begin{gather}
\alpha_{a_2}(t) = 0.8 \left( t - m_{a_2}^2 \right) + 2,\\
\alpha_{K_2^*}(t) = 0.83 \left( t - m_{K_2^*}^2 \right) + 2 ,
\end{gather}
with $\alpha_M^{\prime}$ denoting the slope of the trajectory $\alpha_M(t)$.

\section{NUMERICAL RESULTS}
\label{sec:level3}
Using the formalism established in Sec.~\ref{sec:level2},  we calculate the cross sections for the production of the predicted pseudotensor mesons in the specific reactions $\pi^-p\to\eta_2n$ and $K^-p\to\eta_2\Lambda$ on a proton target. For a generic $2 \to 2$ process, the differential cross section in the center of mass frame is given by:
\begin{equation}
\frac{d\sigma}{dt} = \frac{1}{64\pi s} \frac{1}{|{p}_{1\mathrm{cm}}|^2} \overline{|\mathcal{M}|^2},
\end{equation}
where $s = (p_1 + p_2)^2$ is the square of the center of mass energy, ${p}_{1\mathrm{cm}}$ is the momentum of incident pion or kaon in this frame, and $\overline{|\mathcal{M}|^2}$ denotes the squared scattering amplitude, averaged over initial spins and summed over final spins.

\subsection{Cross section for $\pi^{-} p\to\pi_2^- (1670) p$ reaction}

To validate our framework, we first calibrate it against the well-measured reaction $\pi^{-} p\to\pi_2^- (1670) p$, which involves a known pseudotensor meson. We optimize the cut-off parameter $\Lambda$ by fitting the experimental total cross section data. The analysis employs nine data points from Ref.~\cite{Ascoli:1973nj} spanning a beam momentum range of $P_{\text{Lab}} = 3.1 - 30.0$ GeV/$c$, minimizing the $\chi^2$ per degree of freedom (dof). Since the branching fraction of $\pi_2(1670) \to f_2(1270)\pi$ was suggested to be 56.3\% by the PDG, the coupling constant $g_{\pi_2(1670)f_2(1270)\pi} \simeq 5.439\ \mathrm{GeV}$ is obtained with $\Gamma_{\pi_2 \to f_2\pi} \simeq 0.145\ \mathrm{GeV}$. The Regge trajectory for the exchanged $f_2$ meson is taken as $\alpha_{f_2}(t) = 0.9 ( t - m_{f_2}^2 ) + 2$~\cite{Irving:1977ea}. The resulting optimal cut-off is $\Lambda = 0.762$ GeV, yielding $\chi^2/\text{dof} = 1.006$, indicating good agreement with the experimental data.

\begin{figure}[h!]
    \centering
    \includegraphics[width=0.45  \textwidth]{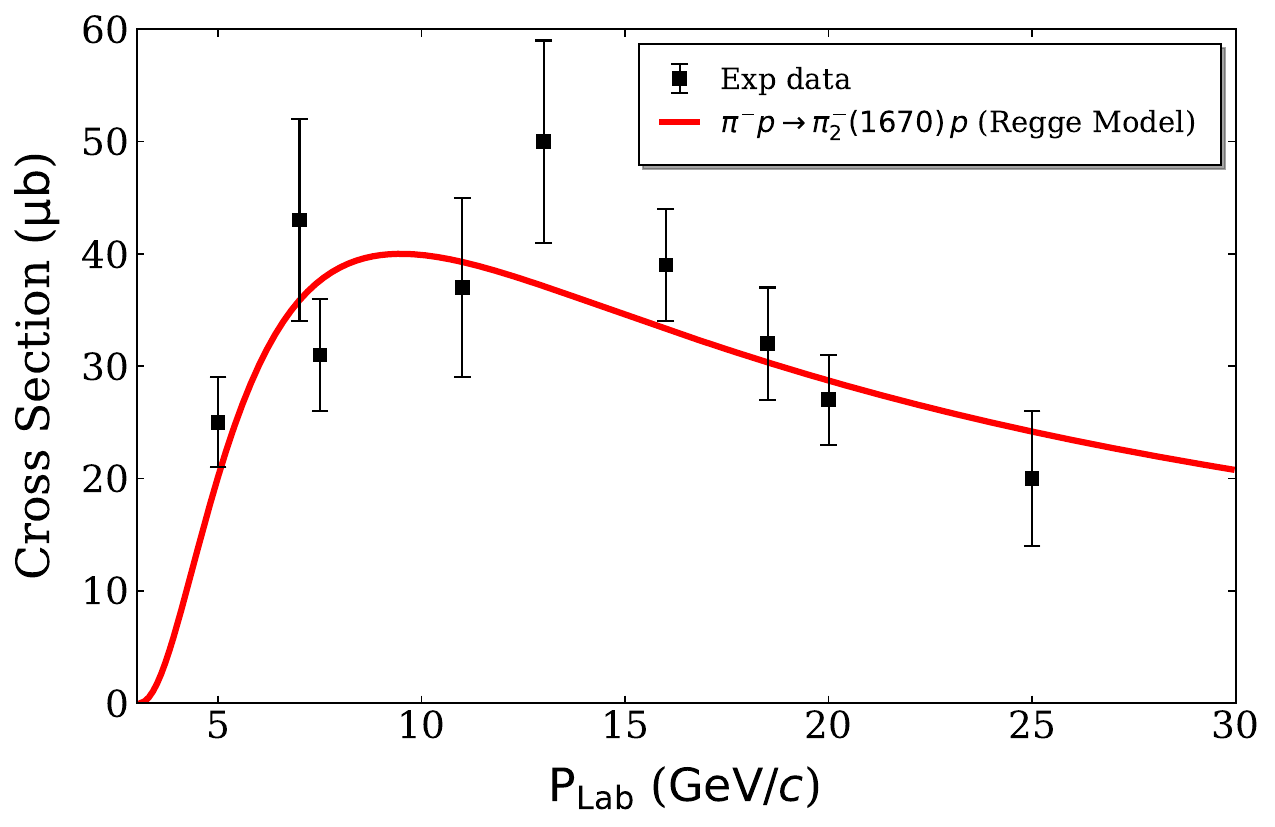} 
    \caption{\label{fig:2}Total cross section for the $\pi^{-} p\to\pi_2^- (1670) p$ reaction. The red solid line represents the result of our Regge model calculation. The experimental data are taken from Ref.~\cite{Ascoli:1973nj}}
\end{figure}

Figure~\ref{fig:2} shows the measured cross section for the reaction $\pi^- p \rightarrow \pi_2^-(1670) p$ as a function of beam momentum. As shown, the Regge theoretical prediction (red solid line) accurately describes the data. The cross section rises sharply from the reaction threshold, reaches a peak of approximately $40.0~\mu\mathrm{b}$ around $9.5~\mathrm{GeV}/c$, and subsequently decreases slowly with increasing beam momentum. 
At lower momenta, specifically at $5~\mathrm{GeV}/c$, the theoretical prediction of $20~\mu\mathrm{b}$  is slightly below the measured value of
$25\pm 4~\mu\mathrm{b}$ , though still consistent within the experimental uncertainty and the expected theoretical systematic errors.
For higher momenta ($P_{\mathrm{Lab}}>16$ GeV/$c$), the model agrees well with all data points within uncertainties. In particular, the prediction of $24~\mu\mathrm{b}$ at 25 GeV/$c$ is consistent with the measured value of $20\pm6~\mu\mathrm{b}$. This successful description validates the applicability of our Regge framework at high energies.

\subsection{Cross section for $\pi^{-} p\to\eta_2n$ reactions}

\begin{figure}[h]
    \centering
    \includegraphics[width=0.48  \textwidth]{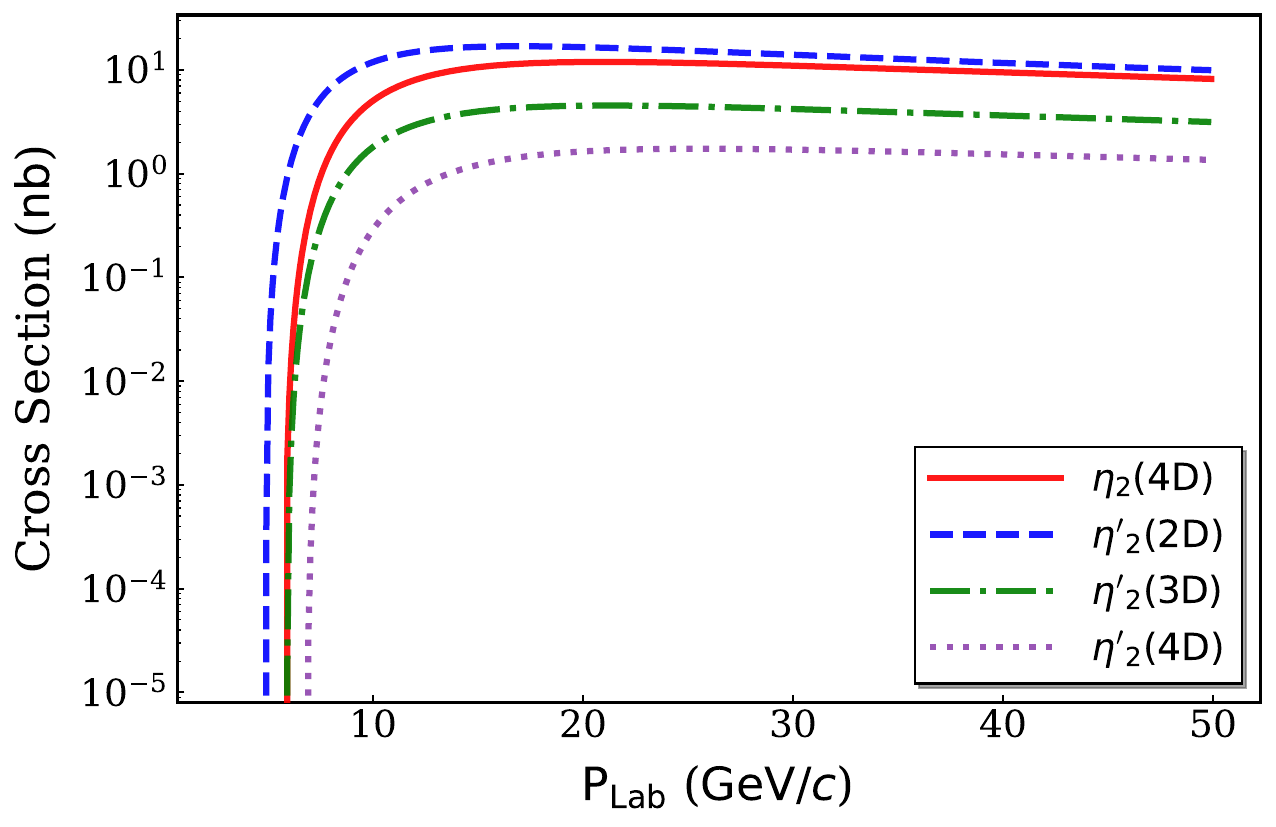} 
    \caption{\label{fig:3}Total cross section for the $\pi^{-} p\to\eta_2n$ reaction in the Regge model.}
\end{figure}

We now present the calculated cross sections for the pion induced production of the predicted isoscalar pseudotensor mesons. Fig.~\ref{fig:3} shows the total cross sections for $\pi^{-} p \to \eta_2 n$ as a function of the pion laboratory momentum $P_{\text{Lab}}$. All four states exhibit a characteristic behavior: a rapid rise near threshold due to the opening of the production channel, followed by a gradual decrease at higher momenta, consistent with the typical energy dependence of Reggeized $t$-channel exchanges.

The four predicted pseudotensor meson states exhibit distinct production characteristics, with their thresholds and peak cross sections occurring at specific pion beam momenta, as detailed in Table~\ref{tab:production_characteristics}. A clear correlation exists between the mass of these states and their production requirements: higher-mass pseudotensor mesons require higher beam momenta to achieve their peak cross sections.

\begin{table}[h]
\centering
\caption{Production characteristics in pion induced reactions.}
\label{tab:production_characteristics}
\renewcommand{\arraystretch}{1.2}
\begin{tabular*}{\linewidth}{@{\extracolsep{\fill}} lcccc}
\hline
& \multicolumn{4}{c}{Pseudotensor State} \\
\cline{2-5}
Parameter & $\eta_2(\mathrm{4D})$ & $\eta_2^{'}(\mathrm{2D})$ & $\eta_2^{'}(\mathrm{3D})$ & $\eta_2^{'}(\mathrm{4D})$ \\
\hline
Production Threshold (GeV/c) & 5.9 & 4.9 & 5.9 & 6.9 \\
Peak Cross Section (nb) & 11.95 & 16.91 & 4.54 & 1.74 \\
$P_{\mathrm{Lab}}$ at Peak (GeV/c) & 20.9 & 16.9 & 21.3 & 25.6 \\
\hline
\end{tabular*}
\end{table}

Specifically, the $\eta'_2(\mathrm{2D})$ state yields the largest cross section, while the $\eta'_2(\mathrm{4D})$ state attains the smallest. Notably, both the $\eta_2(\mathrm{4D})$ and $\eta'_2(\mathrm{3D})$ states reach their maximum at close beam momenta; however, the former exhibits a significantly larger cross section than the latter at this momentum, implying distinct production mechanisms. For experimental searches of these states via pion-proton scattering, the $P_{\mathrm{Lab}}$ values corresponding to the cross section peaks provide suitable momentum windows.

\begin{figure}[h]
    \centering
    \includegraphics[width=0.48  \textwidth]{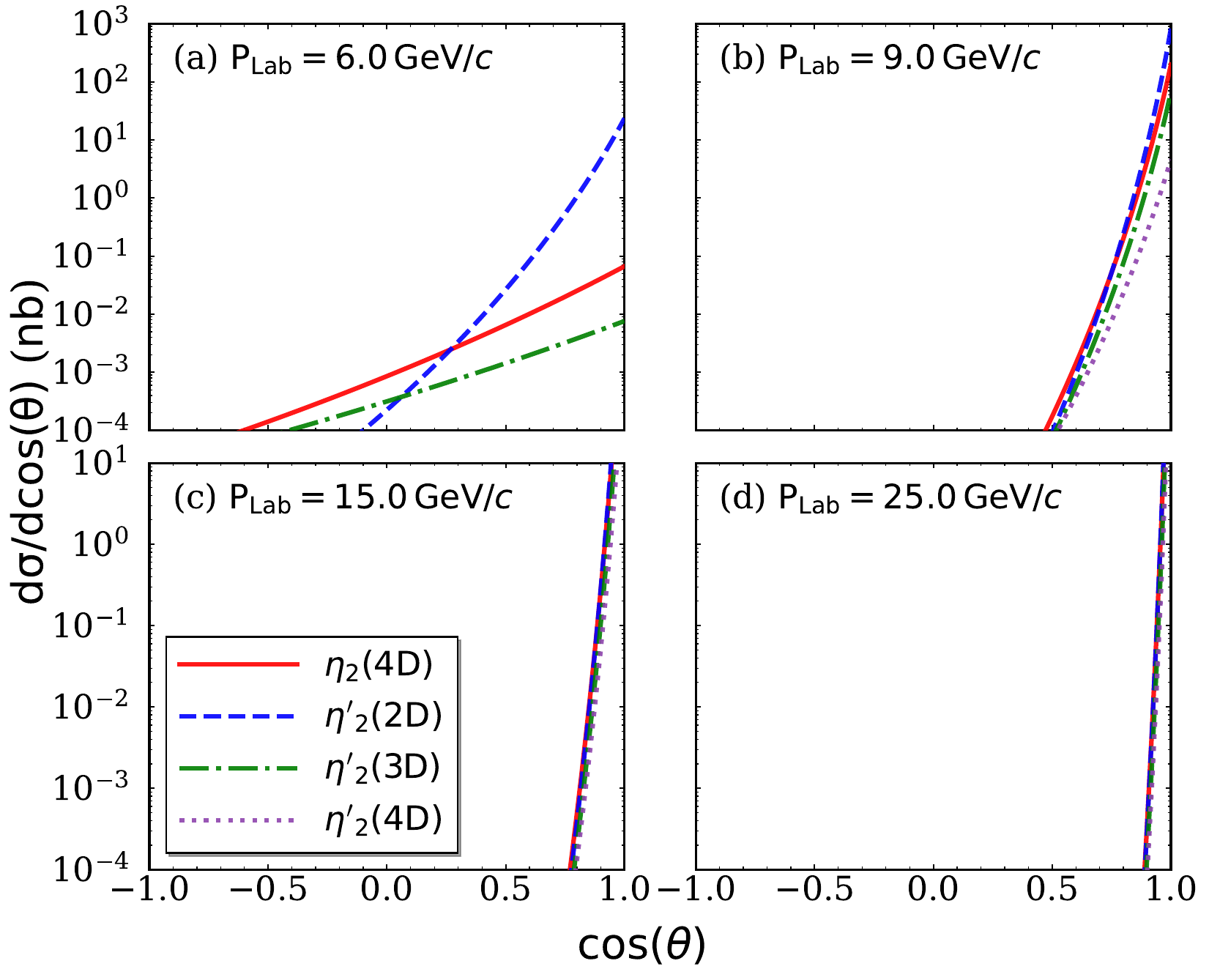} 
    \caption{\label{fig:4}The differential cross section $d\sigma/d\cos\theta$ of the $\eta_2(\mathrm{4D})$ , $\eta_2^{'}(\mathrm{2D})$, $\eta_2^{'}(\mathrm{3D})$, and $\eta_2^{'}(\mathrm{4D})$  production for the $\pi^{-} p\to\eta_2n$ reaction at different $P_{\text{Lab}}$ energies.}
\end{figure}

In addition to the total cross section, the predicted differential cross section for the $\pi^{-} p \to \eta_2 n$ reaction, calculated using the Regge trajectory model, is presented in Fig.~\ref{fig:4}. The results reveal a pronounced angular dependence on the scattering angle $\theta$. As the energy increases, the reaction exhibits a progressively enhanced forward-peaking behavior, indicative of dominant Reggeized $t$-channel exchange processes. This energy-dependent forward peak suggests that experimental verification would be most effectively achieved through measurements at small scattering angles.

\subsection{Cross section for $K^- p \to\eta_2 \Lambda$ reactions}

\begin{figure}[h]
    \centering
    \includegraphics[width=0.48  \textwidth]{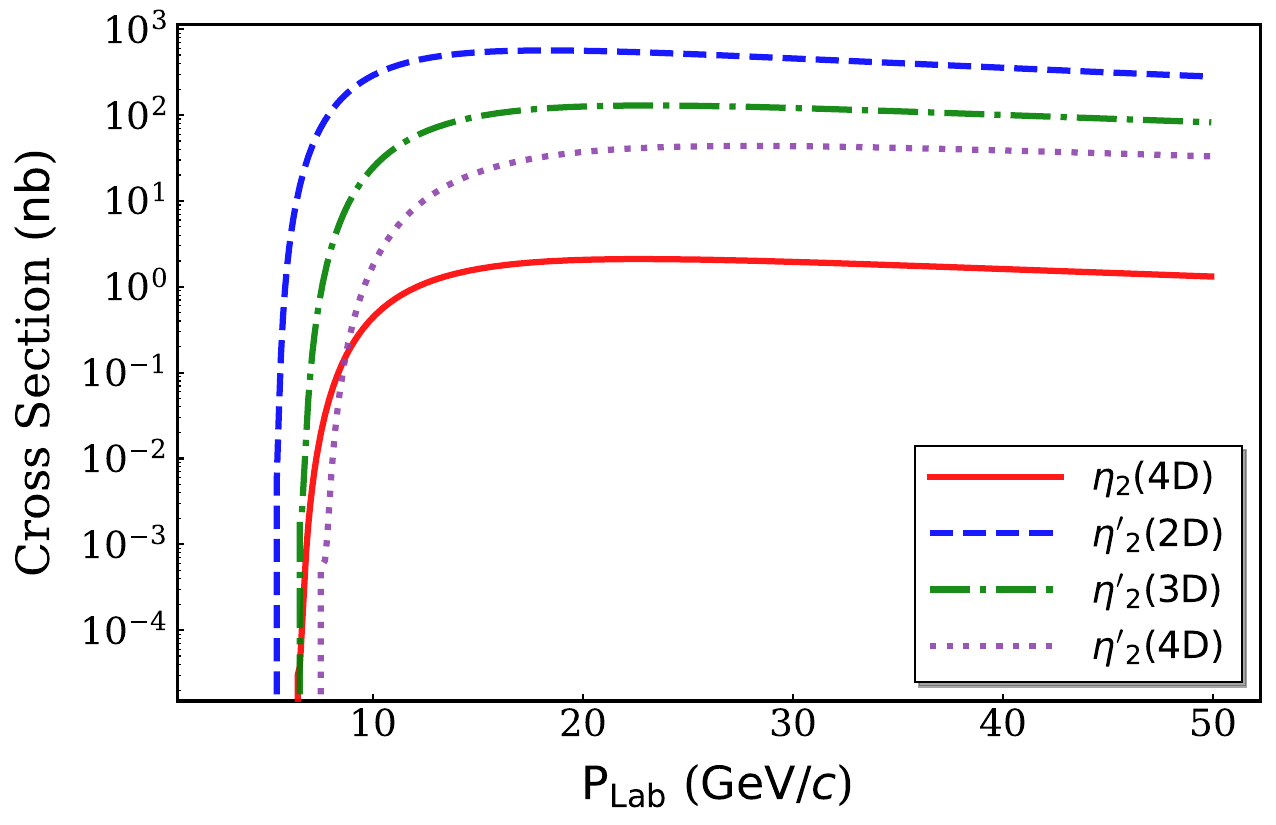} 
    \caption{\label{fig:5}Total cross section for the $K^- p \to\eta_2 \Lambda$ reaction in the Regge model.}
\end{figure}

The predicted production cross sections for the kaon induced reactions $K^- p \to \eta_2 \Lambda$ are presented in Fig.~\ref{fig:5}. Although the functional dependence on $P_{\text{Lab}}$ is similar to that of pion induced reactions, the cross section magnitudes differ significantly. Specifically, the cross sections for the $\eta_2^{\prime}(nD)$ states are substantially larger in kaon induced processes. This enhancement is attributed to their stronger coupling to the $K{K_2}^*$ channel compared to the $\pi a_2$ channel, as evident from the coupling constants listed in Table~\ref{tab:eta2_states_properties}. Conversely, the $\eta_2(4D)$ state exhibits the opposite trend, with a larger production cross section in pion induced reactions, consistent with its relatively stronger coupling to $\pi a_2$.

Table~\ref{tab:kaon_production} summarizes the thresholds and peak cross sections that characterize the production of the four predicted pseudotensor meson states in kaon induced reactions. A clear hierarchy is observed: higher-mass states generally require greater beam momenta to be produced and to reach their peak cross sections.

\begin{table}[h]
\centering
\caption{Production characteristics in kaon induced reactions.}
\label{tab:kaon_production}
\renewcommand{\arraystretch}{1.2}
\begin{tabular*}{\linewidth}{@{\extracolsep{\fill}} lcccc}
\hline
& \multicolumn{4}{c}{Pseudotensor State} \\
\cline{2-5}
Parameter & $\eta_2(\mathrm{4D})$ & $\eta_2^{\prime}(\mathrm{2D})$ & $\eta_2^{\prime}(\mathrm{3D})$ & $\eta_2^{\prime}(\mathrm{4D})$ \\
\hline
Production Threshold (GeV/$c$) & 6.4 & 5.4 & 6.5 & 7.5 \\
Peak Cross Section (nb) & 2.11 & 566.92 & 129.39 & 43.70 \\
$P_{\mathrm{Lab}}$ at Peak (GeV/$c$) & 22.7 & 18.2 & 23.1 & 27.9 \\
\hline
\end{tabular*}
\end{table}

Notably, the $\eta_2(4D)$ state yields a peak cross section that is significantly smaller than those of the other states. The beam momenta at which the cross sections peak define the optimal kinematic regions for detecting these states in future kaon beam experiments.

\begin{figure}[h]
    \centering
    \includegraphics[width=0.48  \textwidth]{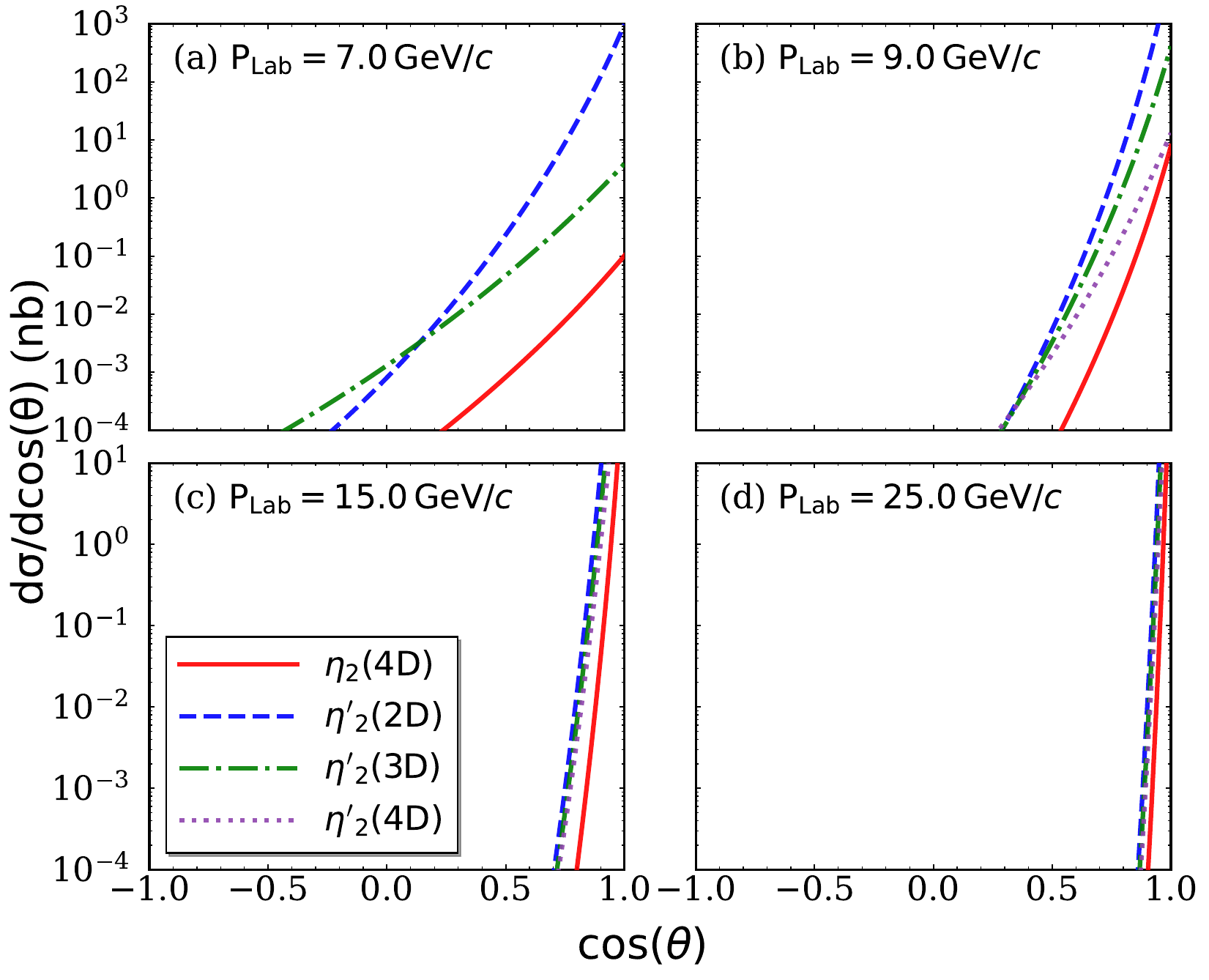} 
    \caption{\label{fig:6}The differential cross section $d\sigma/d\cos\theta$ of the $\eta_2(\mathrm{4D})$ , $\eta_2^{'}(\mathrm{2D})$, $\eta_2^{'}(\mathrm{3D})$, and $\eta_2^{'}(\mathrm{4D})$  production for the $K^- p \to\eta_2 \Lambda$ reaction at different $P_{\text{Lab}}$ energies.}
\end{figure}

Furthermore, the predicted differential cross section for the $K^- p \to \eta_2 \Lambda$ reaction, calculated within the Regge trajectory framework, is presented in Fig.~\ref{fig:6}.  Consistent with the $\pi^{-} p \to \eta_2 n$ case, the differential cross section shows a strong angular dependence on the scattering angle $\theta$, with substantial contributions at forward angles that increase with beam momentum.

As a theoretical preliminary investigation, the phenomenological model based on effective Lagrangians and Reggeization inevitably carries theoretical uncertainties. The primary sources of these uncertainties are the effective Lagrangian methodology itself, the parametrization of the Regge trajectories, and the SU(3) flavor symmetry assumptions applied to the coupling constants. Although these uncertainties influence the predicted absolute magnitudes of the cross sections, they do not compromise the qualitative production patterns nor the relative strengths between different reaction channels established in this study.

\section{\label{sec:level4}SUMMARY}

We have systematically investigated the production mechanisms of four predicted isoscalar pseudotensor mesons -- $\eta_2(\text{4D})$, $\eta'_2(\text{2D})$, $\eta'_2(\text{3D})$, and $\eta'_2(\text{4D})$ -- in pion and kaon induced reactions. Employing an effective Lagrangian approach with Regge trajectory phenomenology, we calculated both total and differential cross sections for the processes $\pi^{-} p \to \eta_2 n$ and $K^{-} p \to \eta_2 \Lambda$.

To validate our theoretical framework, we first calibrated the model parameters using existing data for the related $\pi^{-} p \to \pi_{2}^{-}(1670) p$ reaction, successfully reproducing its cross section within the Regge formalism. This established the reliability of our approach for predicting production cross sections of the unobserved states.

Our calculations reveal several key findings:
\begin{itemize}
    \item A clear production selectivity emerges due to the differing coupling strengths: the $\eta_2(4D)$ state, with its stronger coupling to $\pi a_2$ is significantly more prominent in pion induced reactions, whereas the $\eta_2^{\prime}$ states, with their stronger couplings to $KK_2^*$, are strongly enhanced in kaon induced processes.
    \item All states exhibit characteristic peaks in their total cross sections at specific beam momenta, defining optimal kinematic windows for experimental detection.
    \item The differential cross sections show pronounced forward-angle enhancements, consistent with dominant $t$-channel Reggeized exchange mechanisms.
\end{itemize}

This work provides the first comprehensive theoretical predictions for production cross sections of these mesons, offering crucial guidance for future experimental searches at facilities such as J-PARC, COMPASS, and SPS@CERN. The distinct production patterns and kinematic dependencies identified here establish concrete pathways for discovering and characterizing these states in forthcoming meson-beam experiments.

\begin{acknowledgments}
This work is supported by  the National Natural Science Foundation of China under Grant No.~12405104, the Natural Science Foundation of Hebei Province under Grant No.~A2022203026, the Higher Education Science and Technology Program of Hebei Province under Contract No.~BJK2024176, and the Research and Cultivation Project of Yanshan University under Contract No.~2023LGQN010.
\end{acknowledgments}

\nocite{*}

\bibliographystyle{apsrev4-2}
\bibliography{apssamp}

\end{document}